\documentclass{article}
\usepackage{amsmath}

\input{tcilatex}

\begin{document}

\title{Truth and Completeness in Quantum Mechanics: A Semantic Viewpoint}
\author{Claudio Garola \\
%EndAName
Dipartimento di Fisica e Sezione INFN - Universit\`{a} di Lecce}
\maketitle

\begin{abstract}
The Einstein, Podolski and Rosen (EPR) argument aiming to prove the
incompleteness of quantum mechanics (QM) was opposed by most EPR's
contemporary physicists and is not accepted within the standard
interpretation of QM, which maintains that QM is a complete theory. An
analysis of the semantic implications of the opponent positions shows that
they imply different notions of truth. The introduction of a nonclassical
notion of truth within the standard interpretation is usually justified by
referring to known theorems that should prove that QM is a contextual and
nonlocal theory. However, these theorems are based on a doubtful implicit
epistemological assumption. If one renounces it, one can provide an
alternative interpretation of QM that it realistic in a semantic sense.
Within this interpretation the EPR viewpoint is recovered and QM is
considered a (semantically) incomplete, noncontextual and local theory.
Furthermore, the new interpretation provides several suggestions for
constructing a more general theory embedding QM and for connecting QM with
classical physics and relativity.

\bigskip

\end{abstract}

\section{Introduction}

The famous paper by Einstein, Podolski and Rosen (EPR) aiming to prove the
incompleteness of quantum mechanics (Einstein \textit{et al}., 1935) opened
a debate that is still alive among scholars concerned with the foundations
and the philosophy of QM. It is well known that the EPR position was not
accepted by the majority of contemporary physicists, and that the viewpoint
of EPR's opponents (above all, Bohr's) became the official doctrine of
quantum physicists (briefly, even if imprecisely, \textit{Copenhagen}, or 
\textit{standard}, interpretation).

Basing on several papers by ourselves on topics connected with the EPR
issue, we want to show in this article that a careful analysis of the
semantic implications of the two positions greatly helps to understand the
terms of the debate and provides a new perspective for solving old quantum
problems and avoiding known quantum paradoxes. To this end, we provide a
brief summary of the EPR's argument in Sec. 2, and resume their opponents'
arguments in Sec. 3. Then we argue in Sec. 4 that EPR's viewpoint is
compatible with the adoption of a classical (Tarskian) conception of truth
as correspondence for a suitable fragment of the observative language of QM,
while the standard interpretation adopts a non-Tarskian verificationist
conception of truth for the same fragment. The choice of the standard
interpretation is commented on in Sec. 5, where we remind that it is usually
supported by some famous theorems that are maintained to prove that QM is
necessarily a contextual and nonlocal theory. In the same section we show,
however, that these theorems are based on an implicit epistemological
assumption which contradicts the claimed anti-metaphysical character of QM.
Thus, if one renounces this assumption, the standard interpretation does not
appear any more as logically necessary, and one can envisage a new
interpretation of QM that adopts the classical conception of truth mentioned
above, so that, according to it, QM would turn out to be a noncontextual,
local and incomplete theory, consistently with the EPR philosophical
position. Finally, we remind in Sec. 6 that an interpretation of this kind
has been actually constructed by ourselves (semantic realism, or SR,
interpretation) and observe that this interpretation provides a number of
hints for constructing a more general theory embedding QM and for connecting
QM with classical physics (CM) and relativity. In particular, the SR
interpretation suggests that the difficulties encountered by physicists when
trying to interpret CM as a limit of QM, or to join consistently QM and
relativity, could depend on the adoption of different concepts of truth
within these physical theories (quantum truth, which produces contextuality
and nonlocality, within the standard interpretation of QM; classical truth,
which produces noncontextuality and locality, within CM and relativity).
This incommensurability of theories would then be avoided if the SR
interpretation of QM is accepted.

\section{The EPR incompleteness argument}

Let us schematize briefly the essentials of the EPR argument by using an
updated terminology and following the scheme provided by Jammer in his
fundamental book on the philosophy of QM (Jammer, 1974).

The EPR paper can be divided into four parts, as follows.

Part 1. The following conditions are stated.

\textit{Condition of completeness: ``...every element of the physical
reality must have a counterpart in the physical theory''.}

\textit{Condition of reality: ``If, without in any way disturbing a system,
we can predict with certainty (i.e., with probability equal to unity) the
value of a physical quantity, then there exists an element of physical
reality corresponding to this physical quantity''.}

Part 2. There are pairs of (noncompatible) observables whose values cannot
both be known with arbitrary precision in QM (the position and the momentum
of a particle are observables of this kind). Hence, two alternatives occur.

(i) The values of the two observables are not simultaneously real.

(ii) QM is an incomplete theory.

Part 3. By considering a ``combined'' physical system consisting of two
physical systems (actually, particles), say I and II, it is possible to
prepare the combined system in a state S in which a measurement of the
position of system I would allow one to predict with certainty, by using the
laws of QM, the value of the position of system II, while a measurement of
the momentum of system I would allow one to predict, again by using the laws
of QM, the value of the momentum of system II. Furthermore, ``since at the
time of measurement the two systems no longer interact, no real change can
take place in the second system in consequence of anything that may be done
in the first system''. Hence, both the position and the momentum of system
II correspond to elements of physical reality whenever the combined system
is in the state S.

Part 4. The conclusion in Part 3 shows that the alternative (i) in Part 2
must be rejected. Hence, QM is an incomplete theory.

It is interesting to note that EPR's argument is based on counterfactual
reasonings, since the measurements of position and momentum on system I
cannot be carried out simultaneously. The authors are well aware of this,
for they wrote at the end of the paper:

\begin{quotation}
``... one would not arrive at our conclusion if one insisted that two or
more physical quantities can be regarded as simultaneous elements of reality 
\textit{only when they can be simultaneously measured or predicted}. On this
point of view, since either one or the other, but not both simultaneously,
of the quantities $P$ and $Q$ can be predicted, they are not simultaneously
real. This makes the reality of $P$ and $Q$ depend upon the process of
measurement carried out on the first system, which does not disturb the
second system in any way. No reasonable definition of reality could be
expected to permit this.''
\end{quotation}

Thus, EPR based the defence of their conclusion on a sharp distinction
between reality and empirical knowledge of reality. This distinction
anticipates on the ontological ground the distinction between truth and
epistemic accessibility of truth that will be discussed in Sec. 5.

\section{The objections to the EPR argument}

The reactions to the incompleteness argument summarized in Sec. 1 were
immediate and generally critical (we refer again to Jammer's book quoted in
Sec. 2 for a detailed report on this issue). In particular, Bohr's rejection
of EPR's conclusions (Bohr, 1935a, 1935b) was mainly based on his
`relational conception of quantum states'. According to this conception,
particles and experimental arrangements form an inseparable unit, so that no
independent element of reality can be attributed to system II (hence EPR's
criterion of reality is refuted). The two measurements considered in the EPR
argument are essentially different, and the results obtained in them cannot
be attributed to system II, separating this system from the two different
``wholeness'' to which it belongs in the two cases. Therefore, QM is a
complete theory in the sense that there is nothing in the world besides what
QM allows us to describe.

Bohr's position resumed above is not completely accepted in the current
manuals (see, \textit{e.g}., Messiah, 1961; Cohen-Tannoudij \textit{et al}.,
1973; Greiner, 1989) reporting the standard interpretation of QM, where a
position closer to Heisenberg's (Heisenberg, 1961) is adopted. Indeed,
whenever one considers a physical system made up by two subsystems, one
usually accepts considering the properties of one of the component
subsystems. Yet, for every state $S$ of the whole system one distinguishes
between properties that are certainly possessed by the subsystem (i.e.,
properties that are \textit{real}, or \textit{actual}, in $S$) and
properties that may be possessed or not (i.e., properties that are \textit{%
potential} in $S$). In the physical situation described by EPR some
properties of system II (to be precise, the property of having a sharp
position and the property of having a sharp momentum), that are not actual
in the state $S$, are actualized by the measuring processes, which also
change the state. Yet, the property actualized in a position measurement is
different from the property actualized in a momentum measurement, and the
two properties do not refer to the same final state nor can be attributed to
system II simultaneously. Thus, again, QM must be considered a complete
theory in the sense expounded above.

The foregoing criticisms share a common feature. Both, indeeed, consider QM
as a \textit{contextual} (in the sense that the measurement context cannot
be separated by the physical system that is inquired) and \textit{nonlocal}
(in the sense that contextuality holds also at a distance) theory. Of
course, nonlocality implies contextuality and contradicts EPR's \textit{%
locality} assumption that a measurement on system I cannot induce real
changes on system II if the two systems no longer interact (Sec. 2).

Though nonlocality be fascinating because of the olistic perspective that it
introduces in physics, its consequences are upsetting. Let us consider
indeed its most immediate implications.

At a physical level it emerges an \textit{action at a distance} that is
completely different from the physical interactions described by the
formalism of classical and quantum mechanics.

At an ontological level, some potential (real) properties of a physical
system become real (potential) because of choices of an observer who can be
as far away as desired from the physical system that is considered (though
the observer cannot decide at will the final results of his measurements).

At a semantic level, there are statements that attribute physical properties
to the system but have no truth value (hence no meaning) until some new
knowledge about the system is attained by a far away observer.

Nothwithstanding the problems pointed out above, nonlocality is accepted by
most physicists as an intrinsic and unavoidable feature of QM. We will try
to explain in Sec. 5 why an interpretation of QM has been adopted that is
highly problematical, and to discuss whether there are alternatives to it.
To this end, however, an intermediate step is needed.

\section{Truth and semantic completeness in QM}

The two different viewpoints about the completeness of QM discussed in Secs.
2 and 3 can be better understood in our opinion if one momentarily leaves
apart any philosophical committment and concentrates on a semantic
investigation about the language of QM, pointing out in particular the truth
criteria underlying it according to the two perspectives and the
consequences about completeness (now meant in a purely semantic sense) that
follow from these truth criteria. An investigation of this kind has been
done by ourselves in some recent papers (Garola and Sozzo, 2004; Garola,
2005) and we cannot report it here in details. The essentials of our
arguments can, however, be resumed as follows.

First of all, we must remind some features of the general epistemological
perspective, or \textit{received viewpoint} (Braithwaite, 1953; Hempel,
1965) that we adopt in this paper. According to this perspective, any
physical theory $\mathcal{T}$ is stated by means of a general language that
contains, in particular, a \textit{theoretical language} (which constitutes
the formal apparatus of $\mathcal{T}$ and contains terms denoting
theoretical entities) and an \textit{observative language }$^{1}$. The
former is linked to the latter by means of \textit{correspondence rules},
which provide a partial and indirect interpretation of it. The latter is
interpreted by means of \textit{assignment rules}, which make some symbols
of the observative language correspond to macroscopic entities, as preparing
or registering devices, outcomes of measurements, and so on.

Basing on the above scheme, let us observe now that, via assignment rules,
any interpretation of $\mathcal{T}$ adopts, often implicitly, a \textit{%
theory of truth}, which defines truth values for some (not necessarily all)
statements of the observative language. Whenever an interpretation is given,
we call $\mathcal{T}$ \textit{semantically objective }with respect to a
fragment $\mathcal{L}$ of the observative language if and only if the theory
of truth adopted in it defines truth values for all elementary statements of 
$\mathcal{L}$. Furthermore, we call $\mathcal{T}$ \textit{semantically
complete }with respect to $\mathcal{L}$ if and only if, in any given
physical situation, it allows to predict the truth values of all statements
of $\mathcal{L}$ that have a truth value according to the truth theory that
has been adopted.

Let us come to EPR. These authors adopted in their paper an ontological form
of realism (\textit{local realism}, see Secs. 2 and 3) which is compatible
with a classical theory of truth as correspondence, as explicated rigorously
by Tarski (Tarski, 1956, 1944), for the observative language of QM. To be
precise, let us consider an elementary statement $E(x)$ that attributes a
physical property $E$ to a sample $x$ of the physical system in a given
state $S$ (briefly called \textit{physical object} in the following). This
statement belongs to the standard observative language of QM, and the set of
all statements of this kind constitutes a fragment of the observative
language. Let us denote by $\mathcal{E}$ the set of all properties that the
physical object $x$ can possess, and let us denote the aforesaid fragment by 
$\mathcal{E}(x)$, so that, obviously, $\mathcal{E}(x)\mathcal{=}\{E(x)\mid
E\in \mathcal{E}\}$. Then EPR's local realism is compatible with the
adoption of a classical (Tarskian) theory of truth for $\mathcal{E}(x)$ $%
^{2} $ (hence for a predicate calculus with standard connectives which has $%
\mathcal{E}(x)$ as set of elementary formulas; we do not insist on this
issue for the sake of brevity). According to this theory all statements of $%
\mathcal{E}(x)$ have simultaneous truth values (\textit{true}/\textit{false}%
), hence QM is semantically objective with respect to $\mathcal{E}(x)$. Yet,
since the truth values of all statements of $\mathcal{E}(x)$ cannot be
simultaneously predicted by QM, QM is not semantically complete with respect
to $\mathcal{E}(x)$ $^{3}$.

It is now important to remind that, according to the classical (Tarskian)
theory of truth, truth values are defined by means of a set-theoretical
model, so that their definition is independent of the existence of
procedures which may lead to know them. Briefly, \textit{truth} and \textit{%
epistemic accessibility\ of truth} are different concepts. Thus, it can
occur that the truth value of a statement $E(x)$ is defined but cannot be
known in a given physical situation.

Let us come to the standard interpretation of QM. This implies that a
quantum concept of truth exists which is basically different from the
classical concept. Indeed, the standard interpretation adopts a
verificationist position which abrogates the distinction between truth and
epistemic accessibility of truth. To be precise, it assumes that a statement
of the form $E(x)$ has a truth value if and only if it is possible to
provide an empirical proof of it, thus unifying definition of truth and
empirical accessibility of truth. The concept of empirical proof implied by
this definition, however, is not trivial. Indeed, according to QM, every
measurement on a physical object $x$ generally modifies the state of $x$, so
that the obtained result refers to the final state and cannot be taken as
indicative of the truth value of $E(x)$. Taking into account this basic
feature, and the canonical discussions usually carried out in order to
provide an experimental basis to the indeterminacy relations (Jammer, 1974;
Messiah, 1961; Cohen-Tannoudij \textit{et al}., 1973; Greiner, 1989) one
concludes that the following criterion of truth is implicitly adopted within
the standard interpretation of QM.

\textbf{EV} (\textit{empirical verificationism}) $^{4}$. A statement of the
form $E(x)$ has a truth value (\textit{true}/\textit{false}), hence a
meaning, whenever the physical object $x$ is in the state $S$, if and only
if an empirical test can be carried out that specifies this truth value
without altering the state $S$ of $x$.

In order to distinguish the concept of truth introduced by the standard
interpretation of QM from the classical (Tarskian) concept of truth, we say
that $E(x)$ is Q-true (Q-false) in the following whenever $E(x)$ is true
(false) in the sense established by criterion EV.

Criterion EV has some relevant consequences. Indeed, it implies that,
whenever $x$ is given, hence its state $S$ is known, not all statements of
the fragment $\mathcal{E}(x)$ have a truth value, so that QM is not
semantically objective with respect to $\mathcal{E}(x)$. Moreover, $\mathcal{%
E}(x)$ can be partitioned into three subsets, as follows.

$\mathcal{E}_{S}(x)$ : the set of all statements of $\mathcal{E}(x)$ that
are Q-true. One can prove that this set coincides with the set of all
statements that are Q-true for every physical object in the state $S$.

$\mathcal{E}_{S}^{\bot }(x)$ : the set of all statements of $\mathcal{E}(x)$
that are Q-false. One can prove that this set coincides with the set of all
statements that are Q-false for every physical object in the state $S$.

$\mathcal{E}_{S}^{I}(x)$ : the set $\mathcal{E}(x)\setminus \mathcal{E}%
_{S}(x)\cup \mathcal{E}_{S}^{\bot }(x)$ of all statements of $\mathcal{E}(x)$
that are meaningless, or \textit{indeterminate}.

It follows from the definitions above that the set $\mathcal{E}_{S}(x)\cup 
\mathcal{E}_{S}^{\bot }(x)$ is the set of all statements of $\mathcal{E}(x)$
that have a quantum truth value whenever the physical object $x$ is in the
state $S$. \textit{A priori}, the knowledge of $S$ does not imply the
knowledge of $\mathcal{E}_{S}(x)\cup \mathcal{E}_{S}^{\bot }(x)$. It can be
proved, however, that this implication holds in QM. Hence, QM allows one to
deduce the truth values of all statements of $\mathcal{E}_{S}(x)\cup 
\mathcal{E}_{S}^{\bot }(x)$, whenever $\mathcal{S}$ is given. This means
that QM can predict the truth values of all statements of $\mathcal{E}(x)$
that have a truth value according to criterion EV. Thus, we conclude that QM
is semantically complete with respect to $\mathcal{E}(x)$.

For the sake of brevity we understand the reference to the fragment $%
\mathcal{E}(x)$ in the following. Thus, we can comment on the results
obtained above by saying that any conclusion about the semantic objectivity
and the semantic completeness of QM is strictly linked to the theory of
truth that one adopts for the language of QM (of course, this illustrates a
general feature of physical theories: a physical theory that is semantically
objective, or complete, whenever a given theory of truth is chosen, may be
semantically nonobjective, or incomplete, if a different choice is done).

\section{Back to EPR}

We have seen in Sec. 4 that two different theories of truth for (a fragment
of) the observative language of QM underlie the EPR argument and the
standard interpretation of QM. It is then apparent that the choice of a
Tarskian theory of truth would imply a local and noncontextual
interpretation of QM because of semantic objectivity (we have already noted
in Sec. 4 that the Tarskian definition of truth does not make any reference
to the procedures that can be used in order to know truth values, in
particular measurement procedures). This interpretation would avoid a bulk
of problems (in particular, those following from nonlocality and pointed out
at the end of Sec. 3). One is thus led to wonder why physicists have refused
this seemingly plain alternative.

The answer to the foregoing question can be mainly found in the existence of
a number of `no-go theorems', the most famous of which are the
Bell-Kochen-Specker (briefly Bell-KS) theorem (Bell, 1966; Kochen and
Specker, 1967; Mermin, 1993), which is maintained to prove the contextuality
of QM, and the Bell theorem (Bell, 1964; Mermin, 1993), which is maintained
to prove the nonlocality of QM. Both, of course, imply semantic
nonobjectivity of QM $^{5}$. Because of these theorems, any attempt to
provide a noncontextual and local interpretation of QM is usually rejected
with annoyance by quantum physicists, and considered a na\"{\i}ve
misunderstanding of the subtelties of quantum physics.

Nevertheless, the physical meaning of the no-go theorems can be disputed. A
critical analysis of it has been carried out by ourselves together with
other authors (Garola and Solombrino, 1996a, 1996b; Garola, 1999, 2000,
2002, 2003; Garola and Pykacz, 2004) and it is relevant in our opinion since
it opens the way to a new interpretation of QM which adopts a Tarskian
theory of truth, hence it is noncontextual and local, in good agreement with
EPR's position. Therefore, let us briefly summarize it.

First of all, let us preliminarily remind that, according to the received
viewpoint (Sec. 4), every physical theory $\mathcal{T}$ states a number of 
\textit{theoretical physical laws} by using its theoretical language. These
laws have no direct empirical interpretation, but one can deduce from them,
via correspondence rules, \textit{empirical physical laws}, which are
expressed by means of the observative language of $\mathcal{T}$ and then
associated, via assignment rules, with empirical procedures that allow one
to confirm or falsify them.

Let us come now to QM. In order to apply the above scheme to this theory, we
must refine the last part of it, since empirical laws cannot be checked in
every physical situation according to QM. Indeed, there are physical
situations that can be described within the metalanguage of QM but are not 
\textit{empirically accessible}, in the sense that it is impossible, in
principle, to produce empirically a situation of this kind or recognize
empirically whether it occurs. If one wonders about the validity of
empirical laws within these physical situations, one can choose between two
different epistemological positions, that are inspired to ontological
realism (OR) and empiricism (E), and that can be loosely synthetized as
follows.

\textbf{OR}. The mathematical apparatus of $\mathcal{T}$ mirrors some kind
of physical reality, and theoretical laws parallel (up to some unavoidable
approximations) the laws of physical reality. It follows that the empirical
laws deduced from theoretical laws are valid in every \textit{conceivable}
physical situation, be it empirically accessible or not.

The above assumption about the validity of empirical physical laws has been
called \textit{metatheoretical classical principle}, or MCP, in some
previous papers (Garola and Solombrino, 1996a, 1996b; Garola, 1999; Garola
and Pykacz, 2004).

\textbf{E}. The mathematical apparatus of $\mathcal{T}$ is considered as a
rational construction that is justified by its ability of connecting and
predicting empirical facts. Theoretical laws have no truth value, since they
are not directly interpreted, and the empirical laws deduced from them are
valid in all \textit{accessible} physical situations, \textit{i.e}., in
those physical situations in which they can be confirmed or falsified (while
they can be valid as well as not valid in nonaccessible physical situations).

The above assumption about the validity of empirical physical laws has been
called \textit{metatheoretical generalized principle}, or MGP, in some
previous papers (Garola and Solombrino, 1996a, 1996b; Garola, 1999, 2000,
2002; Garola and Pykacz, 2004) $^{6}$.

It is rather odd that standard QM, whose verificationist attitude (Sec. 4)
is explicitly rooted into antimetaphysical committments, adopts implicitly
position OR when deducing the no-go theorems mentioned above. In order to
justify this demanding statement, let us look into this topic in more
details by referring to a specific sample case, that is, the Bell-KS theorem
(Garola and Solombrino, 1996b; Garola, 2002; Garola and Pykacz, 2004).

The Bell-KS theorem is always proved \textit{ab absurdo} in the literature.
One firstly introduces an objectivity condition (O), as follows.

\textbf{O}. All physical observables have simultaneous values in all
physical situations independently of our knowledge of them.

Then, one introduces the condition that the values of physical observables
are consistent with quantum laws (KS condition) as follows.

\textbf{KS}. If $f(A,B,C,...)=0$ is an empirical law of QM (where $A$, $B$, $%
C$, ... denote compatible observables), the values $a$, $b$, $c$ ... of $A$, 
$B$, $C$, ..., respectively, are such that $f(a,b,c,...)=0$.

Finally, one shows that assuming conditions O and KS together leads to a
contradiction. Since condition KS seems to follow directly from QM, hence it
cannot be denied without renouncing QM, one deduces that assumption O is
untenable within QM. But assumption O is obviously equivalent to the
assumption that QM is semantically objective. It follows that QM is a
semantically nonobjective, hence contextual, theory (which prohibits the
adoption of a Tarskian theory of truth).

The above proof seems conclusive. Nevertheless, if one looks more deeply
into it, one sees that the repeated application of the KS condition produces
physical situations that are not empirically accessible. Indeed, each
empirical law that is applied contains only mutually compatible observables,
but there are observables in some laws that are not compatible with
observables that appear in other laws. The physical situation in which all
these observables have simultaneous values can be conceived but cannot be
empirically recognized, hence it is not empirically accessible in the sense
specified above. Applying repeatedly the KS condition implies assuming that
empirical quantum laws are valid in a situation of this kind, which is
legitimate only if one assumes implicitly position OR about the validity of
the empirical laws of MQ.

Our analysis has far-reaching consequences. Indeed, it implies that
condition KS does not follow directly from QM, since it requires an implicit
epistemological assumption about the range of validity of empirical laws to
QM. Hence, the contradiction proved by the Bell-KS theorem can be avoided
not only renouncing condition O and maintaining condition KS, but also
maintaining O and renouncing the validity of empirical laws of QM within
nonaccessible physical situations, which amounts to limit the validity of
condition KS and say that it cannot, generally, be applied repeatedly (note
that this does not invalidate any standard result in QM) $^{7}$. It follows
that the semantic nonobjectivity of QM is not a logical necessity. Moreover,
a similar analysis can be done by referring to the Bell theorem, showing
that also nonlocality is not a logical necessity (Garola and Solombrino,
1996b; Garola and Pykacz, 2004). One concludes that, contrary to an almost
universal belief, it is possible to envisage new interpretations of QM
adopting a Tarskian theory of truth for the observative language of QM. As
we have anticipated at the beginning of this section, an interpretation of
this kind would avoid the difficulties following from contextuality and
nonlocality and would be compatible with EPR's philosophical position (in
particular, it would classify QM as semantically incomplete). The price to
pay for this, of course, is accepting a more modest gnoseological role of
our physical theories.

We add that the reasonings leading to the no-go theorems (whose mathematical
correctness is undisputable) would still be relevant within the new
interpretation. Indeed, consider again the specific case of the Bell-KS
theorem. The contradiction following from assuming conditions O and KS
together would not show that QM is nonobjective, but that condition KS
cannot be applied repeatedly, hence there must be empirical quantum laws
that cannot hold within nonaccessible physical situations. This suggests
that a physical theory could exist which generalizes QM, reducing to it in
all empirically accessible physical situations only.

\section{The SR interpretation}

An interpretation of QM of the kind envisaged at the end of Sec. 5 has been
actually produced by ourselves, together with other authors, in a series of
papers (Garola, 1991, 1999, 2000; Garola and Solombrino, 1996a, 1996b). It
has been called \textit{SR interpretation}, where SR stands for \textit{%
semantic realism}, since according to it QM is semantically objective, which
is compatible with various forms of realism (in particular, with macroscopic
realism). The SR interpretation has also been supported by some
set-theoretical models that prove its consistency (Garola, 2002, 2003;
Garola and Pykacz, 2004).

Of course, the SR interpretation exhibits all features mentioned at the end
of Sec. 5, since it adopts a Tarskian truth theory for the observative
language of QM. Thus, in some sense, it continues and integrates EPR's
approach, at least from a semantic viewpoint. We maintain that it allows one
to solve many problems, avoiding some well known paradoxes of standard QM.
In particular, the following results can be attained.

(i) No EPR-like paradox occurs, since QM is a local and noncontextual theory
(Garola and Solombrino, 1996b; Garola, 1999).

(ii) The quantum measurement problem is reconsidered and solved in
semiclassical terms (Garola and Pykacz, 2004).

(iii) Some suggestions are provided for embedding QM within the more general
theory mentioned at the end of Sec. 5.

In addition, our arguments in Sec. 5 lead us to suspect that the
difficulties encountered by physicists attempting to attain CM as a limit of
QM, or to unify QM and relativity, could have a common cause. Indeed, both
CM and (special or general) relativity adopt a Tarskian theory of truth for
their observative language, while standard QM introduces, as we have seen in
Sec.3, a notion of quantum truth that is not compatible with Tarski's.
Should this conjecture be true, adopting the SR interpretation would
eliminate the source of the aforesaid difficulties.

Our conjecture above sounds however rather abstract. But our analysis in
Sec. 4 allows us to support it with a more detailed description of some
suggestions provided by the adoption of the SR interpretation. Indeed,
noncontextuality and locality of QM within this interpretation are attained
at the expense of accepting the epistemological position E in Sec. 5, which
limits the range of validity of empirical quantum laws (the only empirical
consequence of this acceptance should be that in any quantum measurement
there are physical objects that are not detected, see Garola, 2003, and
Garola and Pykacz, 2004). Analogous limits could then be assumed for quantum
field theories in order to recover noncontextuality, which is a crucial
issue, since contextuality is a major obstacle to the unification of QM and
relativity. Furthermore, since the SR interpretation is an example of the
class of interpretations envisaged at the end of Sec. 4, it suggests that a
broader theory embedding QM could be constructed, the laws of which differ
from quantum laws only in nonaccessible physical situations. Thus, it is
rather natural to think that CM should be a limiting case of this broader
theory, which explains the partial failure of the attempts to recover CM as
a limiting case of standard QM.

The foregoing suggestions, of course, only sketch a research program, and no
one can guarantee that the expected results will actually be found. Yet, our
program has the merit, at least, of propounding a new approach to old,
unsolved theoretical problems, based on a deep change of physicists'
epistemological attitude rather than on new mathematical techniques and
assumptions. Since the standard approach has been only partially successful
till now, the new program is worth of attention in our opinion.

We would like to close our paper with a brief remark regarding quantum logic
(QL). Indeed, we want to stress that the introduction of an interpretation
that adopts a classical theory of truth for the observative language of QM
does not mean that QL must be rejected. Rather, it leads to interpret QL as
a logical structure formalizing the properties of a metalinguistic concept
that does not coincide with truth (to be precise, the pragmatic concept of 
\textit{justification} in QM). This interpretation is consistent with the
adoption of an \textit{integrated perspective}, which aims to incorporate
classical and nonclassical logical systems into a unified framework that
preserves both the globality of logic and the classical notion of truth as
correspondence (Garola, 2005).

\bigskip

\bigskip

\bigskip

\bigskip

\bigskip

\bigskip

\bigskip

\bigskip

\bigskip

\bigskip

\bigskip

\bigskip

\bigskip

\bigskip

\bigskip

\bigskip

\textbf{ACKNOWLEDGEMENT}

\smallskip

The author is greatly indebted to Carlo Dalla Pozza, Arcangelo Rossi, Luigi
Solombrino and Sandro Sozzo for reading the manuscript and providing useful
suggestions.

\bigskip

\bigskip

\bigskip

\bigskip

\textbf{NOTES}

\smallskip

$^{1.}$ The possibility of distinguishing between theoretical and
observative language of a physical theory has been often criticized on the
basis of the argument that theory and observation are strictly intertwined.
We have already summarized our position about this problem elsewhere (Garola
and Sozzo, 2004, footnote 3). Here we remind only that we agree that the
choice of the observative language depends on the theory and that also the
observative domain can be seen as theory-laden. Yet, we maintain that
theoretical and observative language can still be distinguished if one
accepts Campbell's principle (Campbell, 1920), according to which the part
of the theory embodied within the observative domain must not depend on the
theoretical structure that one wants to interpret (in order to remind this
principle, the observative language could also be called \textit{%
pre-theoretical language}).

$^{2.}$ The EPR argument can easily be restated in terms of statements of
the form $E(x)$. In the specific case considered by EPR, indeed, $x$ may be
taken to indicate a sample of the combined system consisting of systems I
and II, and statements as ``system I has momentum $p$'' or ``system II has
position $r$'' can be translated into the statements ``$x$ has the property
that the momentum of its component system I has value $p$'', and ``$x$ has
the property that the position of its component system II has value $r$'',
respectively, that have the required form. It is then apparent that EPR's
reality condition and locality assumption are compatible with the adoption
of a Tarskian theory of truth for $\mathcal{E}(x)$ (by the way, we remind
that physical properties are usually identified in QM with pairs $(A,\Delta
) $, where $A$ is a physical observable and $\Delta $ a Borel set on the
real line, see, \textit{e.g.}, Beltrametti and Casinelli, 1981; the
properties that appear in our translations can obviously be written in this
way).

$^{3.}$ Note that `realistic' completions of QM can be contrived that are
nonlocal, as exemplified by Bohm's theory (Bohm, 1952a, 1952b). Also this
theory is compatible with the adoption of a Tarskian truth theory for the
observative language of QM. Yet, a statement which attributes a property to
a physical object cannot be considered elementary according to Bohm's
theory, and may have no syntactic expression in some physical situations.

$^{4.}$ From an empirical viewpoint, criterion EV introduces the requirement
that, before performing a test of the truth value of $E(x)$, one has to
prove that the test does not modify the state $S$ of $x$. The empirical
procedure that one must adopt in order to fulfill this requirement is not
obvious, and it must be worked out by taking into account the laws of QM
(Garola, 2005).

$^{5.}$ The assertion that QM is semantically nonobjective can also be based
on theoretical and epistemological considerations (indeterminacy principle
plus a verificationist position) or on empirical arguments (double slit
experiment). Both these approaches, however, are problematical (Garola,
2000), even if the latter is almost universally used in the manuals. The
theorems mentioned above, instead, seem to be conclusive because of their
mathematical rigour.

$^{6.}$ We maintain that MGP could be formalized by formalizing firstly the
observative language of QM and then reconsidering the \textit{truth mode }of
empirical laws within the framework of a Kripkian semantics. In short, one
should define a non trivial accessibility relation on the set of all
possible worlds and assume that empirical laws are valid only in those
worlds that are in the accessibility relation with the real world. However,
we will not discuss this topic in the present paper.

$^{7.}$ Of course, one could avoid any contradiction also giving up both
condition O and condition KS. We do not take into account this possibility
here, since it does not seem plausible from a physical viewpoint.

\bigskip

\bigskip

\textbf{BIBLIOGRAPHY}

\smallskip

BELL\ (John S.), 1964, ``On the Einstein-Podolski-Rosen paradox,'' \textit{%
Physics }\textbf{1}, 195-200.

BELL (John S.), 1966, ``On the problem of hidden variables in quantum
mechanics,'' \textit{Rev. Mod. Phys.} \textbf{38}, 447-452.

BELTRAMETTI (Enrico) and CASSINELLI (Gianni), 1981, \textit{The Logic of
Quantum Mechanics}, Reading (MA), Addison -Wesley.

BOHM (David), 1952a, ``A suggested interpretation of the quantum theory in
terms of `hidden variables,' Part I,'' \textit{Phys. Rev., }\textbf{85},
166-179.

BOHM (David), 1952b, ``A suggested interpretation of the quantum theory in
terms of `hidden variables,' Part I,'' \textit{Phys. Rev., }\textbf{85},
180-193.

BOHR (Niels), 1935a, ``Quantum mechanics and physical reality,'' \textit{%
Nature} \textbf{136}, 65.

BOHR (Niels), 1935b, ``Can quantum-mechanical description of physical
reality be considered complete?'' \textit{Phys. Rev.} \textbf{48}, 696-702.

BRAITHWAITE (Richard B.), 1953, \textit{Scientific Explanation}, Cambridge,
Cambridge University Press.

CAMPBELL (Norman R.), 1920, \textit{Physics: The Elements}, Cambridge,
Cambridge University Press.

COHEN-TANNOUDJI\ (Claude), DIU (Bernard) and LALO\"{E} (Frank), 1973, 
\textit{Mecanique Quantique}, Paris, Hermann.

EINSTEIN (Albert), PODOLSKI (Boris) and ROSEN (Nathan), 1935, ``Can quantum
mechanical description of physical reality be considered complete?'' \textit{%
Phys. Rev.} \textbf{47}, 777-780.

GAROLA (Claudio), 1991, ``Classical foundations of quantum logic,'' \textit{%
Int. J. Theor. Phys.} \textbf{30}, 1-52.

GAROLA (Claudio), 1999, ``Against `paradoxes': A new quantum philosophy for
quantum physics,'' in \textit{Quantum Physics and the Nature of Reality}, D.
Aerts and J. Pykacz, eds., Dordrecht, Kluwer.

GAROLA (Claudio), 2000, ``Objectivity versus nonobjectivity in quantum
mechanics,'' \textit{Found. Phys.} \textbf{30}, 1539-1565.

GAROLA (Claudio), 2002, ``A simple model for an objective interpretation of
quantum mechanics,'' \textit{Found. Phys.} \textbf{32}, 1597-1615.

GAROLA (Claudio), 2003, ``Embedding quantum mechanics into an objective
framework,'' \textit{Found. Phys. Lett.} \textbf{16}, 605-612.

GAROLA (Claudio), 2005, ``A pragmatic interpretation of quantum logic,'' 
\textit{quant-ph/0507122}.

GAROLA (Claudio) and PYKACZ (Jaroslaw), 2004, ``Locality and measurements
within the SR model for an objective interpretation of quantum mechanics,'' 
\textit{Found. Phys. }\textbf{34}, 449-475.

GAROLA (Claudio) and SOLOMBRINO (Luigi), 1996a, ``The theoretical apparatus
of semantic realism: A new language for classical and quantum physics,'' 
\textit{Found. Phys. }\textbf{26}, 1121-1164.

GAROLA (Claudio) and SOLOMBRINO (Luigi), 1996b, ``Semantic realism versus
EPR-like paradoxes: The Furry, Bohm-Aharonov and Bell paradoxes'' \textit{%
Found. Phys. }\textbf{26}, 1329-1356.

GAROLA (Claudio) and SOZZO (Sandro), 2004, ``A semantic approach to the
completeness problem in quantum mechanics,'' \textit{Found. Phys. }\textbf{34%
}, 1249-1266.

GREINER (Walter), 1989, \textit{Quantum Mechanics, an Introduction}, Berlin,
Verlag.

HEISENBERG (Werner), 1961, ``Planck's discovery and the philosophical
problems of atomic physics'' in \textit{On Modern Physics}, London, Orion
Press.

HEMPEL (Carl. G.), 1965, \textit{Aspects of Scientific Explanation}, New
York, Free Press.

JAMMER (Max), 1974, \textit{The Philosophy of Quantum Mechanics}, New York,
Wiley.

KOCHEN (Simon) and SPECKER (Ernst P.), 1967, ``The problem of hidden
variables in quantum mechanics,'' \textit{J. Math. Mech.} \textbf{17}, 59-87.

MESSIAH (Albert), 1961, \textit{Quantum Mechanics}, Amsterdam, North Holland.

MERMIN (N. David), 1993, ``Hidden variables and the two theorems of John
Bell,'' \textit{Rev. Mod. Phys. }\textbf{65}, 803-815.

TARSKI (Alfred), 1944, ``The semantic conception of truth and the
foundations of semantics,'' in L. Linski (ed.), \textit{Semantics and the
Philosophy of Language}, Urbana, University of Illinois Press.

TARSKI (Alfred), 1956, ``The concept of truth in formalized languages,'' in
Tarski, A., \textit{Logic, Semantics, Metamathematics}, Oxford, Blackwell.

\vspace{7in}

\textbf{Riassunto.} Gli argomenti introdotti da Einstein, Podolski e Rosen
(EPR) per dimostrare l'incompletezza della meccanica quantistica (MQ) furono
respinti dalla maggioranza dei fisici contemporanei di EPR e non sono
accettati nell'inter-pretazione standard della MQ, secondo cui la MQ \`{e}
una teoria completa. Se si analizzano le implicazioni semantiche delle due
posizioni in conflitto si deduce che esse sottendono due diverse nozioni di
verit\`{a}. L'introduzione di una nozione non classica da parte
dell'interpretazione standard \`{e} usualmente giustificata facendo
riferimento ai noti teoremi che proverebbero che la MQ \`{e} necessariamente
una teoria contestuale e non locale. Comunque, questi teoremi sono basati su
una dubbia assunzione epistemologica implicita. Se tale assunzione viene
evitata, \`{e} possibile concepire un'interpretazione alternativa
all'interpretazione standard che \`{e} realistica in senso semantico.
Nell'ambito di questa interpretazione \`{e} possibile recuperare il punto di
vista di EPR, e la MQ \`{e} considerata una teoria (semanticamente)
incompleta, locale e non contestuale. Inoltre la nuova interpretazione
fornisce alcuni suggerimenti per costruire una teoria pi\`{u} generale che
incorpori la MQ e per connettere la MQ con la fisica classica e la relativit%
\`{a}.

\bigskip

\bigskip

\bigskip

\bigskip

\bigskip

\bigskip

\bigskip

\bigskip

\bigskip

\textbf{Draft.} Completed on October 26th, 2005.

\textbf{Author.} Claudio Garola.

\textbf{Biographic notes.} Born in Asti (Italy) on May 3rd, 1941. Degree in
Physics at the University of Turin (1967). Formely assistant-professor in
General Physics (1970), then associated professor in General Physics (1985)
and in Theoretical Physics (1988) at the University of Lecce. At present,
full professor of Logic and Philosophy of Science at the same University.

\textbf{Research interests.} General Physics, Algebra, Theoretical Physics,
Foundations of Quantum Mechanics, Foundations of Physics, Quantum Logic.

\textbf{Private address.} Via Antonaci 4,

73100 Lecce, Italy.

\textbf{Institution address.} Dipartimento di Fisica e Sezione INFN,

Universit\`{a} di Lecce,

Via Arnesano,

73100 Lecce, Italy.

\textbf{Phone numbers.} (+39) 0832 297438 (Dept.); (+39) 0832 603302 (home);
(+39) 0832 359655 (home).

\textbf{E-mail.} garola@le.infn.it

\end{document}